\documentclass[aps,prl,superscriptaddress,reprint,showpacs,nofootinbib]%
{revtex4-1}
\usepackage{amsfonts,amsmath}

\DeclareMathOperator{\Tr}{Tr}

\begin{document}
\title{
Qudit-Based Measurement-Device-Independent Quantum Key Distribution Using Linear Optics
}

\author{H. F. Chau}
\thanks{Corresponding author, email: \texttt{hfchau@hku.hk}}
\affiliation{Department of Physics, University of Hong Kong, Pokfulam Road,
 Hong Kong}
\affiliation{Center of Theoretical and Computational Physics, University of
 Hong Kong, Pokfulam Road, Hong Kong}
\author{Cardythy Wong}
\affiliation{Department of Physics, University of Hong Kong, Pokfulam Road,
 Hong Kong}
\author{Qinan Wang}
\affiliation{Department of Physics, University of Hong Kong, Pokfulam Road,
 Hong Kong}
\author{Tieqiao Huang}
\affiliation{Department of Physics, University of Hong Kong, Pokfulam Road,
 Hong Kong}
\date{\today}

\begin{abstract}
 Measurement-device-independent (MDI) method is a way to solve all detector
 side-channel attacks in quantum key distribution (QKD).
 However, very little work has been done on experimentally feasible qudit-based
 MDI-QKD scheme although the famous (qudit-based) round-robin
 differential-phase-shift (RRDPS) scheme is vulnerable to attacks on
 uncharacterized detectors.
 Here we report a mother-of-all QKD protocol on which all provably secure
 qubit-based QKD schemes known to date including the RRDPS and the
 so-called Chau15 schemes are based.
 We also report an experimentally feasible MDI system via optical
 implementation of entanglement swapping based on a recent qudit teleportation
 proposal by Goyal \emph{et al.}
 In this way, we show that all provably secure qudit-based QKD schemes
 discovered to date can be made MDI.
\end{abstract}

\pacs{03.67.Dd, 89.70.-a}

\maketitle

 In quantum key distribution (QKD), two cooperative agents (commonly called
 Alice and Bob) try to share a secret key by sending and measuring signals
 through a quantum channel.
 Realistic experimental apparatus, which is never ideal, posts a serious and
 non-trivial threat to the security of QKD as eavesdropper (commonly called
 Eve) may exploit loopholes due to apparatus imperfections~\cite{RMP09,DLQY16}.
 Even worse, such loopholes, some may yet to be found, could be experimental
 setup specific.
 One way to solve the imperfect detector problem is the so-called
 measurement-device-independent (MDI) method~\cite{LCQ12}, which uses
 teleportation or entanglement swapping technique to close all detector side
 channel loopholes once and for all.
 The beauty of MDI method is that the teleportation or entanglement swapping
 measurement can be performed by a third untrustworthy party (commonly called
 Charlie).
 MDI method is applicable to all prepare-and-measure QKD schemes involving the
 transfer of qubits, qudits and continuous-variable quantum modes as long as
 these schemes can be reduced from certain entanglement-based
 ones~\cite{LCQ12,BP12,cv-MDI}.
 Several experiments have demonstrated the feasibility of qubit-based
 MDI-QKD~\cite{mdi-expt1,mdi-expt2}.
 However, it is not clear how to reliably implement the qudit-based MDI
 proposal in Ref.~\cite{BP12}, which relies on the entanglement swapping
 scheme by Bouda and Bu\v{z}ek~\cite{BB01}, using photonics techniques.
 With the discovery of several promising qudit-based QKD schemes, study of
 their MDI version is no longer a pure academic issue.

 One such scheme is the round-robin differential-phase-shift (RRDPS)
 scheme~\cite{SYK14}, which has attracted a few pioneer
 experiments~\cite{rrdps-expt1,rrdps-expt2,rrdps-expt3,rrdps-expt4}.
 While this scheme is robust against encoding
 errors~\cite{sourcerobustness_rrdps} and does not need to monitor signal
 disturbance~\cite{SYK14}, it is insecure against several detector
 attacks~\cite{mdtrust_rrdps1,mdtrust_rrdps}.
 Moreover, it is not clear if the MDI version of the RRDPS scheme exists.

 Another scheme is the so-called Chau15 scheme~\cite{Chau15}.
 Information is transmitted in this scheme and its extension~\cite{Chau16} via
 preparation and measurement of qubit-like qudits in the form
 $(|j\rangle\pm|k\rangle)/\sqrt{2}$ with $|j\rangle$ and $|k\rangle$ being
 distinct orthonormal states in a $2^n$-dimensional Hilbert space.
 Following Ref.~\cite{LCQ12}, an naive MDI implementation of these scheme is for
 Charlie to perform entanglement swapping by measuring the states Alice and
 Bob send to him along $\{ |j,j'\rangle\pm |k,k'\rangle,
 |j,k'\rangle\pm|k,j'\rangle \}$ for some randomly chosen distinct pairs of
 $(j,k)$ and $(j',k')$.
 However, this naive implementation is insecure as the cheating Charlie may
 project the states Alice and Bob send to him along $|j'\rangle\pm|k'\rangle$
 and $|j\rangle\pm|k\rangle$ respectively before performing the entanglement
 swapping to obtain the phase information of the two states without being
 caught.
 Furthermore, after entanglement swapping, Bob may have to change the value of
 his raw key based on Charlie's measurement result~\cite{LCQ12}.
 This step can only be performed without affecting the key rate if the
 teleportation procedure is compatible with the state preparation procedure of
 Alice and Bob in the sense that the quantum operations needed to transform
 between different teleportation measurement states used (in the qubit case,
 these are the four Pauli operations) can be deterministically mapped to the
 corresponding classical operations acting on Bob's raw key (in the qubit case,
 this is the logical-NOT operation).
 Unfortunately, the teleportation procedure used in Ref.~\cite{BP12} is not
 compatible with the state preparation procedure used in the Chau15 scheme.
 Hence, even if one may optically implement the MDI protocol in
 Ref.~\cite{BP12} in future, it cannot be used make the Chau15 scheme MDI.

 Here we first report a mother-of-all entanglement-distillation-based MDI-QKD
 scheme which can be reduced to all known provably secure qudit-based QKD
 schemes to date.
 Then we show the feasibility of this mother-of-all scheme by reporting a linear
 optics implementation of the required entanglement swapping operation.

\par\medskip\noindent
{\bf The mother-of-all scheme.}
\begin{enumerate}
 \item Let $n>1$, $N\equiv 2^n$ and $GF(N)$ denotes the finite field of $N$
  elements.  Alice and Bob each prepare an entangled state $|\Phi_{00}\rangle
  \equiv \sum_{i\in GF(N)} |i,i\rangle / \sqrt{N}$ and send the second half of
  the state to Charlie through an insecure quantum channel.
 \item Charlie performs entanglement swapping by measuring the states he
  received from Alice and Bob along the basis ${\mathcal B} = \{
  |\Phi_{ab}\rangle \equiv \sum_{i\in GF(N)} (-1)^{\Tr(b i)}
  |i,i+a\rangle/\sqrt{N} \colon a,b\in GF(N) \}$, where $\Tr(i) = i +
  i^2 + i^4 + \cdots + i^{N/2}$ is the absolute trace of $i$.
  Note that all arithmetic in the state ket are done in the finite field
  $GF(N)$.
  (See Ref.~\cite{FF} for an introduction to finite field arithmetic.)
  Charlie publicly announces his measurement result, namely, the values of
  $a,b$ he obtained.
  Bob applies the linear transformation $|i\rangle \mapsto
  (-1)^{\Tr[(a-i)b]} |i-a\rangle$ for all $i\in GF(N)$ to the first
  half of his state.
  (Or equivalently, Alice applies the linear transformation $|i\rangle \mapsto
  (-1)^{-\Tr(i b)} |i+a\rangle$ to the first half of her state.)
  In the absence of Eve and noise, Alice and Bob should now share the
  entangled state $|\Phi_{00}\rangle$.
 \item Alice, Bob and Charlie repeat the above procedure some times to
  accumulate enough entangled states.
  Then Alice and Bob perform channel error estimation, if necessary, plus
  entanglement distillation to get the final almost perfect copies of
  $|\Phi_{00}\rangle$'s.
  They measure their shares of these distilled pairs to get their final key.
\end{enumerate}

 Note that if the state measurement procedure used by Alice and Bob to obtain
 their final key in step~3 above is compatible with the teleportation procedure
 in step~2, we obtain a provably secure qudit-based QKD scheme by the standard
 Shor-Preskill argument~\cite{SP00}.
 More importantly, all provably secure qudit-based QKD schemes to date can be
 deduced from this mother-of-all scheme in this way.
 For instance, if Alice and Bob both project each of their shared distilled
 pairs to states in the form $(|j\rangle\pm|k\rangle)/\sqrt{2}$, we get the
 Chau15 scheme~\cite{Chau15} and its extension~\cite{Chau16}.
 If Alice and Bob project their states in the form $\sum_{i\in GF(N)}
 (-1)^{s_i} |i\rangle/\sqrt{N}$ for $s_i \in GF(2)$ and
 $[|j\rangle\pm|k\rangle]/\sqrt{2}$ respectively, we obtain the RRDPS scheme
 using $N$-dimensional qudits~\cite{SYK14}.
 And if Alice and Bob prepare their states using the method stated in
 Ref.~\cite{Chau05}, we arrive at the so-called Chau05 scheme.
 We state the MDI version of the RRDPS scheme obtained in this way below as
 illustration.

\par\medskip\noindent
{\bf The MDI version of the RRDPS scheme.}
\begin{enumerate}
 \item Alice prepares $\sum_{i\in GF(N)} (-1)^{s_i} |i\rangle / \sqrt{N}$ and
  sends it to Charlie.
  She jots down the values of $s_i$'s.
 \item Bob prepares $[|j\rangle + (-1)^t |k\rangle] / \sqrt{2}$ and sends it to
  Charlie.
  He jots down the values of $t \in GF(2)$ and $j\ne k\in GF(N)$.
  And he uses $t$ as his raw bit.
 \item Charlie jointly measures the states of Alice and Bob along the basis
  ${\mathcal B}$ and announces the state $|\Phi_{ab}\rangle$ he obtains.
 \item Bob announces $j$ and $k$.
 \item Alice uses $\{ s_{k-a} - s_{j-a} - \Tr[b(k-j)] \} \bmod 2$ as
  her raw bit.
 \item Alice and Bob repeat the above steps to get sufficient raw key bits and
  then distill out their final key through error correction and privacy
  amplification.
\end{enumerate}

 Although it is not possible to perform the complete Bell-like measurements in
 step~2 of the mother-of-all scheme using linear optics~\cite{VY99,LCS99}, we
 report a partial implementation below based on Goyal \emph{et al.}'s qudit
 teleportation proposal.
 In Fig.\ 3 of Ref.~\cite{qudit-teleport_optics}, Goyal \emph{et al.} reported
 a way to project a $N$ qudit state to the antisymmetric state $|\Psi\rangle =
 \sum_{P \in S(N)} \varepsilon(P) \prod_{i\in GF(N)} a^\dag_{i,P(i)}
 |\Omega\rangle / \sqrt{N!}$ by means of linear optics and photon number
 resolving detectors.
 Here $S(N)$ is the group of permutations of a set of $N$ elements,
 $\varepsilon(P)$ is the sign of the permutation $P$, $a^\dag_{i,j}$ is the
 creation operator for a photon propagating along path $i$ and orbital angular
 momentum $j$, and $|\Omega\rangle$ is the vacuum state.
 Since $\sum_{P\in S(N)} \varepsilon(P) \langle m| U^{-1} |P(j)\rangle \langle
 m | U^{-1} |P(k)\rangle = 0$ for all $N$-dimensional unitary operator $U$ and
 $j,k,m\in GF(N)$ with $j\ne k$, measuring every qudit of the state
 $|\Psi\rangle$ along the basis $\{ U|i\rangle \colon i\in GF(N) \}$ always
 yields $N$ distinct outcomes.
 In this sense, $|\Psi\rangle$ is the generalization of the singlet state for
 qubits.
 By choosing $U$ to be a direct sum of $N/2$ Hadamard transformations, we have
 the following modified MDI-RRDPS scheme using linear optics.
 (Only those modified steps are shown.)

\par\medskip\noindent
{\bf The MDI version of the RRDPS scheme using linear optics.}
\begin{enumerate}
 \item[2'] Bob randomly group the $N$ elements in $GF(N)$ into $N/2$ pairs in
  the form $\{ (j_i,k_i) \}$.
  He prepares $(N-1)$ distinct qudit states each selected from the basis
  $\bar{\mathcal B} = \{ (|j_i\rangle\pm|k_i\rangle)/\sqrt{2} \}$ and sends
  them to Charlie.
  He jots down the state $[|j\rangle+(-1)^t |k\rangle]/\sqrt{2}$ in
  $\bar{\mathcal B}$ that he has not prepared and uses $t$ as his raw bit.
 \item[3'] Charlie jointly projects the single qudit from Alice and $(N-1)$
  qudits from Bob to $|\Psi\rangle$ and informs Alice and Bob if the projection
  is successful.
 \item[5'] Alice uses $(s_j-s_k) \bmod 2$ as her raw bit.
\end{enumerate}

 Note that the connections between the above two MDI-RRDPS schemes is that the
 $|\Psi\rangle$ used in the latter can be identified with the
 $|\Phi_{0j}\rangle$ of the former for some $j\ne 0$ through the logical
 encoding of each state in the basis set $\bar{\mathcal B}$ by the tensor
 product of the other $N-1$ states in $\bar{\mathcal B}$.
 Since the probability for Charlie to successfully obtain $|\Psi\rangle$ in
 step~3' equals $1/N^2$, the above scheme is practical only when $N$ is small.
 It is instructive to find more efficient way to project a state to
 $|\Phi_{ij}\rangle$.
 Finally, we write down the MDI version of the Chau15 scheme for completeness.

\par\medskip\noindent
{\bf The MDI version of the Chau15 scheme using linear optics.}
\begin{enumerate}
 \item[] Alice, Bob and Charlie follow all the steps in the MDI version of the
  RRDPS scheme using linear optics with the following modifications.
 \item[1''] Alice sends the state $[|j'\rangle+(-1)^s|k'\rangle]/\sqrt{2}$ to
  Charlie.
  She jots down $j'\ne k'\in GF(N)$ and $s\in GF(2)$.
 \item[5''] If $\{j,k\} = \{j',k'\}$, Alice and Bob uses $s$ and $t$ as their
  raw key bits, respectively.
\end{enumerate}

\begin{acknowledgments}
 This work is supported by the RGC grant~17304716 of the Hong Kong SAR
 Government.
\end{acknowledgments}

\bibliographystyle{apsrev4-1}

\bibliography{qc70.2}

\end{document}